\documentclass[conference]{IEEEtran}
\IEEEoverridecommandlockouts
\usepackage{cite}
\usepackage{amsmath,amssymb,amsfonts}
\usepackage{algorithmic}
\usepackage{graphicx}
\usepackage{soul}
\usepackage{wrapfig}
\usepackage{textcomp}
\usepackage{xcolor}
\usepackage{enumitem}
\def\BibTeX{{\rm B\kern-.05em{\sc i\kern-.025em b}\kern-.08em
    T\kern-.1667em\lower.7ex\hbox{E}\kern-.125emX}}
\begin{document}

\title{Edge Centric Secure Data Sharing with \\ Digital Twins in Smart Ecosystems}
%Edge Centric Secure Data Sharing with Multiple Digital Twins in Smart Ecosystem
%\titlerunning{Edge Centric Secure Data Sharing with Digital Twins}
%Secure Data Sharing with Multiple Digital Twins in Smart Connected Ecosystem
%Edge Centric Secure Data Sharing with Multiple Digital Twins \\ in Smart Connected Ecosystem
%Secure Data Sharing with Multiple Digital Twins using Edge in Smart Connected Ecosystem
%Tag Based Access Control for Multiple Digital Twins
\iffalse
\author{Glen Cathey\inst{1} \and
James Benson\inst{2} \and
Maanak Gupta\inst{1} \and
Ravi Sandhu\inst{2}}
%
\authorrunning{Glen Cathey et al.}
\institute{Dept. of Computer Science,
Tennessee Technological University,\\
Cookeville, Tennessee 38505, USA \and
Dept. of Computer Science, Institute for Cyber Security,\\ University of Texas at San Antonio, TX, USA\\
\email {glcathey42@tntech.edu, james.benson@utsa.edu,  mgupta@tntech.edu, ravi.sandhu@utsa.edu}}
\fi
%\iffalse
\author{\IEEEauthorblockN{Glen Cathey\IEEEauthorrefmark{1}, James Benson\IEEEauthorrefmark{2}, Maanak Gupta\IEEEauthorrefmark{3}, and Ravi Sandhu\IEEEauthorrefmark{4}}
\IEEEauthorblockA{\IEEEauthorrefmark{1}\IEEEauthorrefmark{3}{Dept. of Computer Science},
{Tennessee Technological University},
Cookeville, Tennessee 38505, USA \\\IEEEauthorrefmark{2}\IEEEauthorrefmark{4}Dept. of Computer Science, Institute for Cyber Security, University of Texas at San Antonio, TX, USA\\}
\IEEEauthorrefmark{1}glcathey42@tntech.edu,
\IEEEauthorrefmark{2}james.benson@utsa.edu, 
\IEEEauthorrefmark{3}mgupta@tntech.edu,
\IEEEauthorrefmark{4}ravi.sandhu@utsa.edu}
\iffalse
\author{\IEEEauthorblockN{1\textsuperscript{st} Given Name Surname}
\IEEEauthorblockA{\textit{dept. name of organization (of Aff.)} \\
\textit{name of organization (of Aff.)}\\
City, Country \\
email address or ORCID}
\and
\IEEEauthorblockN{2\textsuperscript{nd} Given Name Surname}
\IEEEauthorblockA{\textit{dept. name of organization (of Aff.)} \\
\textit{name of organization (of Aff.)}\\
City, Country \\
email address or ORCID}
\and
\IEEEauthorblockN{3\textsuperscript{rd} Given Name Surname}
\IEEEauthorblockA{\textit{dept. name of organization (of Aff.)} \\
\textit{name of organization (of Aff.)}\\
City, Country \\
email address or ORCID}
\and
\IEEEauthorblockN{4\textsuperscript{th} Given Name Surname}
\IEEEauthorblockA{\textit{dept. name of organization (of Aff.)} \\
\textit{name of organization (of Aff.)}\\
City, Country \\
email address or ORCID}
\and
\IEEEauthorblockN{5\textsuperscript{th} Given Name Surname}
\IEEEauthorblockA{\textit{dept. name of organization (of Aff.)} \\
\textit{name of organization (of Aff.)}\\
City, Country \\
email address or ORCID}
\and
\IEEEauthorblockN{6\textsuperscript{th} Given Name Surname}
\IEEEauthorblockA{\textit{dept. name of organization (of Aff.)} \\
\textit{name of organization (of Aff.)}\\
City, Country \\
email address or ORCID}
}
\fi
\maketitle

\begin{abstract}
Internet of Things (IoT) is a rapidly growing industry currently being integrated into both consumer and industrial environments on a wide scale. While the technology is available and deployment has a low barrier of entry in future applications, proper security frameworks are still at infancy stage and are being developed to fit varied implementations and device architectures. Further, the need for edge centric mechanisms are critical to offer security in real time smart connected applications with minimal or negligible overhead. 

In this paper, we propose a novel approach of data security by using multiple device shadows (aka digital twins) for a single physical object. These twins are paramount to separate data among different virtual objects based on tags assigned on-the-fly, and are used to limit access to different data points by authorized users/applications only. The novelty of the proposed architecture resides in the attachment of dynamic tags to key-value pairs reported by physical devices in the system. We further examine the advantages of tagging data in a digital twin system, and the performance impacts of the proposed data separation scheme. %For proof-of-concept, we focus on the integration of digital twins and Tag Based Access Control (TBAC) in two applications: smart vehicles and smart manufacturing. 
The proposed solution is deployed at the edge, supporting low latency and real time security mechanisms with minimal overhead, and is light-weight as reflected by captured performance metrics.

%The development of further access control oriented (ACO) architectures in academia is necessary to provide efficient and tested alternatives to current industry solutions. 
%In this paper we examine the current available industry IoT solutions provided by Google, Amazon, Oracle, and Microsoft in relation to Tag Based Access Control (TBAC) and propose an architecture that facilitates tag based subdivision of data. Currently major industry solutions offer Role Based Access Control (RBAC) regarding access to digital twins. The novelty of the proposed architecture resides in the attachment of tags directly to key-value pairs reported by physical devices in the system. This allows for efficient and on-the-fly grouping of like data into subsets which then reside in digital twins. Access control can then be done on these twins therefore minimizing data exposure to authorized users. 

%\keywords{Data Security \and Data Sharing \and Edge Computing \and Smart Connected Systems \and Digital Twins \and Access Control \and Tags}
\end{abstract}


\section{Introduction}
Internet of Things (IoT) is an omnipresent technological field with a focus on ease of use as well as a high degree of device autonomy. However, it presents a host of new issues regarding data security as well as substantively increasing the attack surface of the average end user's device ecosystem. The purpose of these IoT devices is to incorporate `smart' automation into the everyday life of the end user, but just as the function of these devices is automatic, the access control mechanisms are as well. As IoT devices become more prevalent in normal user environments, as well as integrated into our societal infrastructure, this automatic data sharing becomes a potential weak-point in the chain of data security. 

Securing the data handled by smart autonomous devices becomes more important as the ubiquity of connected devices grows. If one home in a neighborhood has an IoT connected thermostat, there is not much incentive for adversaries to develop technologies to exploit potential underlying weaknesses. However, if every other house contains a plethora of smart devices, each of which is continuously gathering and transmitting data, there is a much higher potential gain from compromising these devices\cite{205156}. Reality is becoming more reflective of this hypothetical every day. The amount of smart devices worldwide increased by one billion from 2019 to 2020. The global IoT market is projected to nearly triple between 2020 to 2030, from 8.74 to 25.44 billion devices\cite{holst_2021}. 

With this substantial increase in amount of connected devices, the infection rate of these devices is similarly growing. In 2019, compromised IoT devices made up 16.17\% of all infected devices connected to mobile networks, that number more than doubled to 32.72\% in 2020\cite{onestore}. 
The type of data being secured is also changing as these technologies are adopted across different fields, and therefore so are the consequences of data being compromised. Manufacturing environments become `smarter' every day as they integrate IoT devices to increase efficiency and decrease production time. In the US alone, manufacturing is a 2.3 trillion dollar industry which accounts for 11.39\% of GDP\cite{nam}. Interruption to these processes on a macro scale could very well lead to fiscal losses in the billions. Further, smart internet connected cars are becoming more widespread every year, with 51.1 million being sold in 2019 and a projected 76.3 million to be sold in 2023\cite{wagner_2020}. This far-reaching growth leads to higher quality and convenient cars for consumers, however internet connected cars must have sufficient security protocols in place. The reliability and integrity of smart and autonomous car data is critical when user's lives depend on the vehicle functioning as designed. 

The gravity of these security issues highlight the need for more secure frameworks and practices regarding the handling of data generated by IoT devices and connected ecosystem. One way to accomplish this is the integration of \textbf{digital twins} \cite{digitaltwins} into device control and data acquisition. Digital twins, or device shadows\footnote{While digital twin refers to the whole encapsulation of a physical device in software and device shadow refers to the JSON data structure holding a representation of device state, these terms are very similar and are used interchangeably in this paper.}, are the virtual counterpart to physical objects which introduce a layer of abstraction between higher level control of devices and device specific actuation and sensing methods. These shadows can be used to facilitate separation between the object and cloud services layer (detailed in the background section), as well as enabling separation of IoT data into subsets. Digital twins also lead to more consistent interaction between higher level layers and physical devices. Device state as well as current connection status can always be accessed by higher level layers due to the persistent nature of the twin. It should be noted that digital twins comprise only a portion of the overarching architecture. The separation of device communication into layers within an access control oriented (ACO) architecture (discussed in the related work section) is necessary and present. Authentication and subsequently authorization must happen between discrete layers to ensure the architecture as a whole is secure. 

IoT devices often have multiple state data such as user settings, manufacturer configuration, and operational status. Each of these state data needs to be accessed by different users or at differing frequencies. Usually, there is one to one mapping between physical devices and virtual objects, meaning, one can only associate a single virtual object to a single device and are required to store all sets of device state data in one shadow. As a limiting consequence, all users will have access to the entire shadow and can consequently read and update state data they should not have. Further, data tagging plays a large role in our access control mechanisms and serves as the basis for the separation of data in the architecture. Tags are attached to the data generated in the system and are the attribute on which data is separated. This allows for easy and computationally inexpensive grouping of like data and adds further classification for data type. In this way, tags are the metric on which data is separated, and the digital twins are the receptive containers for that data. Once data has been separated and distributed based on the tags it carries, access control is centered around granting access to individual shadows. The data present in a given shadow is directly related to the tags applied to that data, therefore granting access to individual subsets of data within shadows is a form of Tag Based Access Control (TBAC). 

In this paper, we propose a novel approach of data security by using multiple shadows (digital twins) for one physical object, with the intent of separating data among different virtual objects based on tags assigned on the fly, which are then used to limit access to different data points by any authorized users/applications. The proposed solution is deployed at the edge, supporting low latency and real time security mechanisms with minimal overhead, and is light-weight as discussed in the implementation section. The implementation described in this paper is built on version 5.0 of the MQTT\footnote{https://mqtt.org/} protocol, and therefore communication occurs within topics in a publisher-subscriber model. While the referenced implementation applies access control to these topics, any model must include similar access control techniques with regard to flow of information between layers. For proof-of-concept, we focus on the integration of digital twins and TBAC in two industry applications: smart vehicles and smart manufacturing. We will examine the mechanisms for secure data sharing between digital twins, the advantages of tagging data in a digital twin system, and the performance impacts of the proposed data separation scheme. 
 The key contributions of this paper are as follows:
\begin{itemize}
\item Attachment of tags directly to device state information in order to reduce `distance' between access control mechanisms and device data itself.
\item Dynamic and on-the-fly subdivision of device state at the local edge according to attached tags.
\item Limiting data exposure to authorized entities via subdivision of data in a many-to-one relationship between digital twins and physical devices.
\item Implementation of the proposed architecture to reflect the plausibility and efficiency, together with
brief comparative discussion on performance metrics.
\end{itemize}

The remainder of the paper is structured as follows. Section \ref{sec:background} discusses relevant background such as established access control oriented (ACO) architectures and IoT literature. Section \ref{sec:need-for-edge} demonstrates the necessity of security at the local edge in abstract principle as well as in applications such as intelligent vehicles and smart manufacturing. Existing industry solutions and their limitations are also examined in this section. Section \ref{sec:tbac-proposal} defines the proposed architecture and the mechanisms for the attachment of tags within the context of TBAC. Section \ref{sec:evaluation} presents implementation and associated performance metrics. Finally, section \ref{sec:conclusion} summarizes our work and looks ahead to future work in this field.

\section{Relevant Background}
\label{sec:background}
This section reviews primitive building blocks of cloud and edge assisted smart connected systems. In addition, we will also reflect on relevant literature which has offered some security solutions and approaches for IoT and CPS ecosystems. 

\begin{figure*}[th!]
%\vspace{0.1in}
  \centering
  \includegraphics[width=.7\textwidth]{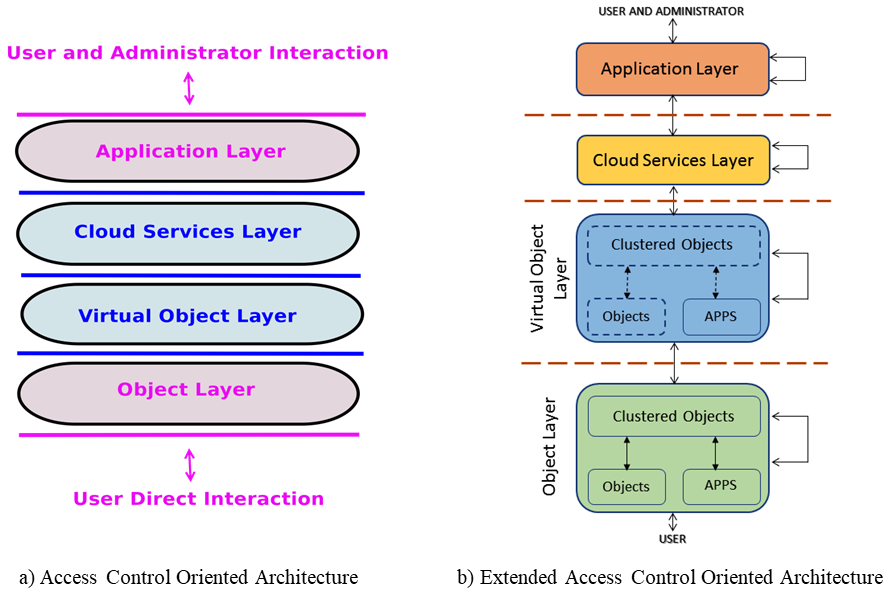}
  \caption{Multi-layered Access Control Focused Architectures}
  \label{Architecture_Layers}
%\vspace{0.1in}
\end{figure*}
\subsection{Access Control Oriented Architectures}
Several access control oriented (ACO) architectures have been proposed in the literature for IoT \cite{7809752,gupta2018authorization,Alshehri2018AccessCM,8673782,weijia2018,weijia2021,userauthIoT,celik_tan_mcdaniel_2019,franch2020,yahyazadeh2019,yahyazadeh2020,gupta2020attribute,dgupta2020access} and cyber physical systems (CPS) such as smart cars \cite{gupta2018authorization}\cite{guptaABAC2019}, intelligent transportation \cite{gupta2020secure} and smart manufacturing \cite{9502070}, which focus on the separation of systems into layers as illustrated in Figure \ref{Architecture_Layers} (a). As shown in
Figure \ref{Architecture_Layers} (b), ACO architecture (proposed by Alsehri and Sandhu\cite{7809752}) has four layers - object, virtual object, cloud services and application – with users and administrators
interacting at object and application layers. In addition, communication can happen
within a layer (shown as self loop in Figure \ref{Architecture_Layers} (b) and the adjacent
layers above and below. It should be noted that the extended access control oriented (E-ACO) \cite{gupta2018authorization} architecture shown in Figure \ref{Architecture_Layers} (b) is an extension to the generic ACO architecture with some additional components as discussed in the following section. 

The \textit{object layer} is comprised of the physical devices which either sense or actuate the environment within which they reside. These devices can be individual or clustered into larger objects (shown in Figure \ref{Architecture_Layers} (b)) which contain many sensors, actuators, etc. There are several examples of clustered objects such as smart cars, mobile phones, or production lines; all of which contain many smart devices connected to a network. The physical objects in the object layer communicate with their digital twins (aka virtual objects) in the virtual object layer. These devices can communicate with other devices using different communication technologies
including Bluetooth, WiFi, Zigbee, LAN, LTE or 5G. Physical devices communicate with their cyber counterparts (virtual objects) using protocols like HTTP, MQTT, DDS or CoAP. Users can also
directly access physical objects at this layer. In an extended access control oriented architecture (E-ACO) as shown in Figure \ref{Architecture_Layers} (b), clustered objects (COs) are introduced, which are objects with multiple sensors, and allow for possible interaction between sensors in same CO or between different object’s sensors. These COs, such as smart cars, also have applications associated with them which offer services to users, in this case drivers. For example, a rear vision system is an application in cars to get rear-view, which gets data from the rear camera (an object) to provide dashboard view to the driver. These applications
in the object layer of E-ACO are add-on's to the object layer in ACO architectures.

The \textit{virtual object} layer holds the digital twins for all of the physical objects in the system. Digital twins in this layer communicate directly with their associated physical objects, the other virtual objects (VOs) present, and the cloud layer. The VOs in this layer hold the last received state of the physical object they represent, as well as processing desired states for those objects. These desired states can be received from other VOs, or the cloud layer. There may also be many virtual objects associated with one physical object. Virtual objects can hold the entire data set generated by their physical object, or subsets of that data. The virtual object layer in E-ACO architecture can have one or many cyber entities (virtual object or digital twins) for both clustered and individual objects. These twins can be created in the cloud layer, or local edge layer to support real time communication. For example, when sensors s$_{1}$ and s$_{2}$  across different clustered objects
communicate with each other, the sequence of communication via
virtual object layer should follow starting s1 to vs$_{1}$  (digital twin of
s$_{1}$), vs$_{1}$  to vs$_{2}$  and vs$_{2}$  to physical sensor s$_{2}$ . 

 The \textit{cloud layer} is the location of long term storage of device state, as well as more complex processing of received device data. Computationally intensive operations can be performed at this layer, thereby easing the burden of devices themselves as well as the hardware at the edge. These operations could include, but are not limited to: image processing with the intent of facial or object recognition, machine learning in order to fine tune a system’s efficiency, or data visualizations. This layer manages communication with the virtual object and application layers, and is responsible for propagating control signals entered by the user as well as generating control signals based on the aforementioned data processing. Communication between clouds can also take place within this layer to enable big-data analytics or the union of discrete but related implementations. Single or multiple cloud scenarios can exist to support
federation or trusted collaboration between them. Some important IoT cloud platforms include Amazon AWS\footnote{https://aws.amazon.com/}, Microsoft
Azure\footnote{https://azure.microsoft.com/en-us/} IoT Hub, and Google Cloud IoT Core\footnote{https://cloud.google.com/iot-core}. An important use for cloud layer in IoT/CPS involves defining security policies for authorized
communication among different objects. 

The \textit{application layer}, is responsible for both displaying system information to the user and for user input. This layer needs to communicate with the layer directly below it to pass on control signals and receive visualizations and system state information. Users and administrators can remotely send commands and instructions to smart devices residing within the bottom layer using these applications, but such
interaction must propagate through the other two ACO middleware layers (cloud services and virtual object)

The layered access control oriented (ACO) structure discussed was proposed by Alsehri and Sandhu\cite{7809752} with a focus on clarifying the middleware layers' function and form in IoT architectures. The distinction between the virtual object layer and the cloud layer lends itself to integration of heterogeneous objects into the system, as well as giving a well defined framework for access control techniques. This work also supports computation at the edge, as opposed to the cloud layer, by delineating the differences therein. Edge computation is necessary in industry with a focus on low latency that necessitate fast response times i.e. autonomous cars, or dynamic agricultural monitoring systems such as drones.

\subsection{Related Work}

Recent extensive analysis of IoT technologies into the field of agriculture has been published by Gupta et al \cite{gupta2020security}. The authors found that computation at the edge is a requirement for many systems with a focus on real time analysis and dynamic behavior. However, the assignment of responsibility at the edge comes with an increased attack surface due to the array of heterogeneous physical devices deployed\cite{sina2020farming}. These devices are usually not designed with security as a chief concern\cite{o2016insecurity}, and are a major security liability if configured incorrectly. The deployment of cryptographic security measures are difficult at the device level due to the computational constraints of most IoT enabled devices. While solutions do exist \cite{dhanda_singh_jindal_2020}, they are relatively novel and have not yet found widespread implementation. They propose a lightweight multi-factor authentication protocol in the form of an independent Certificate Authority (CA). This allows for dynamic authentication and meets the complexity needs at the device level. It is worth noting however, that this solution does not detail practices to limit what data is being shared, only how to grant authorization. 

Another area smart connectivity can greatly improve performance and efficiency is manufacturing. Kusiak\cite{kusiak2018smart} makes the case that due to the trend of ever-increasing integration of smart sensors into manufacturing environments, the utilization of that data will drive further integration of smart actuators and data analysis into manufacturing processes. The employment of this novel data will lead to more accurate and complex modelling, optimization, and simulation. These models will give insight into potential fine-tuning practices to increase manufacturing efficiency, and the analysis of equipment monitoring will lead to predictive maintenance and prevention of equipment failure\cite{s20195480}. This comes at the cost of increased cyber-security and safety concerns. As companies become reliant on modelling and IoT device infrastructure the value of these technologies goes up, therefore their security becomes paramount to continued profit and growth. In regard to safety, as automation and autonomous smart decision-making becomes integral to manufacturing centers, the responsibility of equipment to function correctly continuously shifts to lie upon the cyber-physical implementation.   

\section{Need for Edge Centric Secure Data Sharing}
\label{sec:need-for-edge}

Implementations of IoT technologies at scale involve the generation of large quantities of data, which are used to affect system state by adjusting IoT actuators present in the system. The metrics for this state change are system specific but all systems require the sharing of data generated by local physical devices. This sharing can take place directly from virtual device to virtual device, virtual device to the local edge, or local edge to cloud. Which type of sharing takes place is determined by the level of computation necessary before the system state is affected. 

In all aspects of device data sharing in a smart IoT connected system the local edge is critical and extensively utilized. These edge systems ensure
low latency and real time communication much needed in
most smart applications without bandwidth issues. In such scenarios, the edge plays a role in virtual device to virtual device sharing because all shadow clients in these systems reside on these local edge. Therefore even if the hardware of the edge is unneeded for computations more complex than device hardware can handle, the mechanisms of data sharing between virtual device clients still reside on, and are controlled by the edge which works as a middle man and relay the data. Virtual device to local edge sharing is required to facilitate computations exceeding physical device hardware, aggregation of device data in order to manage the system as a whole, or simply for comprehensive logging of system state. In the case of local edge to cloud data sharing, the local edge acts as a data pass-through in order to supply system information to cloud resources for computations that exceed local edge hardware capacity. These computations may include, but are not limited to, facial or object recognition, complex image processing, or machine learning algorithms. 

Due to the local edge's involvement in all data sharing which take place within an IoT system, the security of edge and the data it holds is of the utmost importance. The architecture proposed in this paper focuses on securing data in the system by managing the allocation of individual pieces of data into dynamic subsets based on tags. This is a form of TBAC with a focus on reducing the `distance' between tags assigned to data and the data itself. The implementations of TBAC currently present at the industry level utilize rules to tag data and independently apply tags to resources. This creates separation between the data and the tags applied to that data, as well as the containers that data will be placed within. We aim to improve this by directly applying tags to data and distributing data into digital twins based on those tags. Therefore each digital twin will have a set of tags defining what subset of data it will hold, and data will be distributed into each twin based on tags attached directly to that data. 
 
 \begin{figure}[t!]
%\vspace{0.1in}
  \centering
  \includegraphics[width=8cm, height=2.8in]{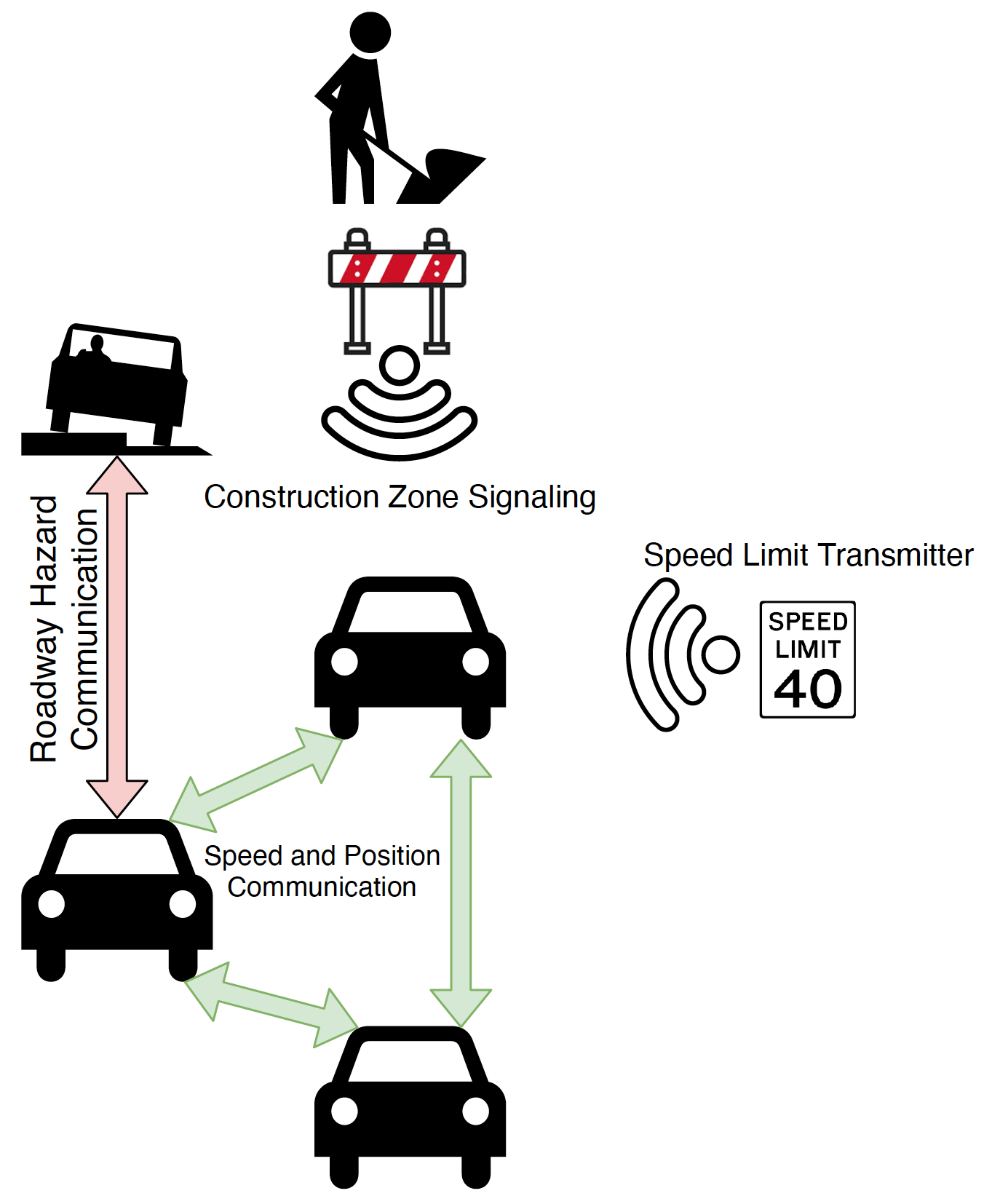}
  \caption{External Smart Car Communication}
  \label{Car_comm}
%\vspace{0.1in}
\end{figure}
\subsection{Motivating Use cases}
\subsubsection{Smart Cars and Intelligent Transportation}
 Smart vehicles require low-latency with high-volume data sharing. The internal network-connected sensors and actuators present in the car must be continuously sharing their data with the edge. This data is processed to allow functionality such as lane assistance systems, emergency collision avoidance, or full autonomous navigation. Externally, the car may be communicating with roadway infrastructure such as traffic lights, speed limit transmitters, or construction zone signalling shown in Figure \ref{Car_comm}. Sharing data with other smart vehicles offers many benefits as well, in the form of automated lane merging protocols, increased speed limits due to increased reliability of surrounding vehicles, and shared awareness of roadway hazards. These factors culminate in smart cars prioritizing internal sensor-to-edge and external edge-to-edge sharing.

 While local and edge-to-edge sharing is prioritized, there is also utilization of the cloud layer in both logging data and implementation of more complex algorithms. User usage data such as location, driving habits, and maintenance history can be stored in the cloud for later retrieval. Performance data generated by the vehicle can also be sent to the cloud for processing by machine learning algorithms in order to monitor system health and send preemptive maintenance alerts. 

\subsubsection{Smart Manufacturing}
 Smart manufacturing environments can take advantage of IoT technologies by distributing large quantities of internet connected smart sensors throughout the production pipeline. The local edge can be used to monitor system health by ensuring that sensor values fall within acceptable operating ranges. The cloud layer ensures system health by employing machine learning algorithms which monitor system efficiencies as reported by sensors in the system and give predictive points of failure. This architecture considers the necessity for low latency response times in the event of critical failure via utilization of the edge as a monitoring system, while also encouraging long-term health of the system via utilization of machine learning resources in the cloud. 

\subsection{Threat Model}
\par The adversary threat model considered in this paper is heavily influenced by the security research put forward by the USDOT Intelligent Transportation Systems Office \footnote{https://www.its.dot.gov/factsheets/cybersecurity.htm}. We have chosen to consider this research in developing our threat model because the environment it studies, smart transportation, is one of the most dynamic and difficult to secure. It is also the most industry applicable environments for IoT requiring edge based solution, as described earlier. The threats and vulnerabilities we address in the proposed solution include:
\noindent
\begin{itemize}[leftmargin=*]
    \item Entities authorized to read or affect system state of objects may get access to extraneous data which they should not have. As an example, roadway infrastructure such as speed limit transmitters should be allowed to affect maximum speed of a smart vehicle, but should not be able to read or write data such as location, personal user data, vehicle specifications, or maintenance information.  In traditional IoT digital twin architectures access is granted as a binary, where users are authorized to view and affect contents of a digital twin as a whole or not at all. This exposes even authorized entities to an excess of data, and is less secure than giving access to individually tagged pieces of data. 
    \item Due to the large number of IoT devices in ecosystem such as smart factories, ITS, or smart homes, it is a near certainty that some of these devices will malfunction. In all of these objects failure may have severe consequences, therefore quick and efficient realization of device malfunction is a necessity. The attachment of tags directly to pieces of data allows for consistent processing and verification regarding the value of that data by the associated digital twin. For example, all values tagged 'temperature' within a system could have bounds implemented as rules such as: temperature should be a positive integer, and temperature should never exceed 100 units. If a piece of data exceeds or falls below these bounds then it is safe to assume that the physical device is malfunctioning and system state is compromised.
\end{itemize}
This paper proposes an edge based solution addressing these security concerns via data distribution into multiple digital twins foundationally built on TBAC. We also support and build upon security properties addressed by USDOT ITS research. We focus on \textbf{Authenticity \& Trust} by implementing open source software such as Mosquitto\footnote{https://mosquitto.org/} which maintains support for multiple forms of authorization including username/password, PSK (Pre-Shared Key), and external plugin support. This allows for system specific authorization schemes to be implemented, while also providing built in authorization methods. \textbf{Confidentiality \& Privacy} is supported in this architecture by the subdivision of data into multiple digital twins. Data exposure is limited by allowing authorized entities to view only the subset of data they require to function, thereby keeping the information in the system confidential and private. %Modification of the system is similarly controlled in order to preserve integrity.  

\subsection{Some Industry Solutions and Limitations}
\subsubsection{Microsoft Azure}
 Microsoft Azure IoT Hub allows attachment of tags to digital twins and physical devices but they are static informational metadata such as device specific location/properties and do not serve a security function nor do they delineate pieces of data. Queries can be used to route data into digital twins based on tags, but are not dynamic and queries must be added to process additional tags. Digital twins in this architecture may not receive subsets of generated data as tags are applied to physical devices, not individual pieces of data. Therefore digital twins may be tagged in order to authorize reception of device data, however this authorization is purely a binary: either they will receive the full device message if tags are matching, or they will receive nothing. This limitation is not present in our implementation because the tags are attached to each key-value pair in every message and therefore messages may be subdivided based on tags.
\subsubsection{Google IoT Core}
Google IoT Core offers a highly scalable industry IoT solution, however does not implement distribution of data based on tags. Tags in their architecture can be applied to physical devices and serve as device identifiers specifying metadata information such as: serial number, location, or manufacturer information. Tags may also be applied at the digital twin level in order to grant access to users authorized to view individual tags. However due to the lack of data distribution based on data present, all data is collected in one digital twin. Therefore subsets of data cannot be accessed and in order to view device data a user must be authorized to view all tags present. This is subversive to the limitation of data sharing in the system and is less dynamic than access control granted to individual shadows and therefore tags. 
\subsubsection{AWS IoT Core}
Amazon Web Services IoT Core supports a many-to-many digital twin-to-physical device relationship in the form of named digital twins accompanying a base unnamed twin. Physical devices may publish data directly to their named shadow counterparts, or publish all data to the base unnamed shadow which can then manage publications to named shadows. The purpose of named shadows is to hold subsets of physical device data in order to minimize data exposure and system malleability upon authorization of a resource regarding access to the shadow. This division of data comes closest to our proposed architecture, however there is no support for tagging discrete pieces of data. Rules can be implemented to distribute data to named shadows, however due to the lack of tag attachment to data these rules must work on data value, associated key, or other system information. This means data can be subdivided in the system, but like data can not be effectively grouped dynamically. Rules must be defined to sort individual data keys into named shadows resulting in a less scalable and more implementation specific system. 
\subsubsection{Oracle IoT Asset Monitoring Cloud Service}
 Oracle’s cloud IoT service allows the creation of digital twins to hold device information, as well as predictive twins to hold the results of complex analysis of device performance such as machine learning and neural networks. They also allow simple creation of rules regarding alerts and system functioning such as location-based rules which activate when a device enters or exits defined locations, threshold-based rules which trigger when a devices reported data either exceeds or falls below set values, and alert-based reactions which trigger physical device actions given alerts present in the system. However the tags which can be attached to devices are purely descriptive and serve no security or access control centric function. Therefore the division of data in this architecture is difficult, as individual pieces of data are not delineated in any way other than their associated keys. Highly dynamic environments may suffer security consequences as authentication in this architecture is a binary of full access or no access. 

\section{Proposed Multiple Digital Twins with \\ Tags Based Access Control}
\label{sec:tbac-proposal}
%Tag Based Access Control
 It is clear at this point that IoT environments generate and subsequently share large amounts of data. Mechanisms for sharing relevant and required information facilitate correct data apportionment between resources, as well as limiting the amount of data shared as much as possible. Minimizing data sharing within the architecture both increases security and decreases the burden on networking hardware. 
 Our approach to controlling data sharing implements subdivision of data generated by physical sensors, and grants individual access to those subsets. This employs the security principle of least privilege by giving access to only the information required by the authorized resource, and allowing system malleability on the smallest surface possible. This increases system security as well as efficiency by minimizing the size of data flowing in the system from producers to consumers. 

\begin{figure*}[t!]
\centering
    \includegraphics[scale=0.51]{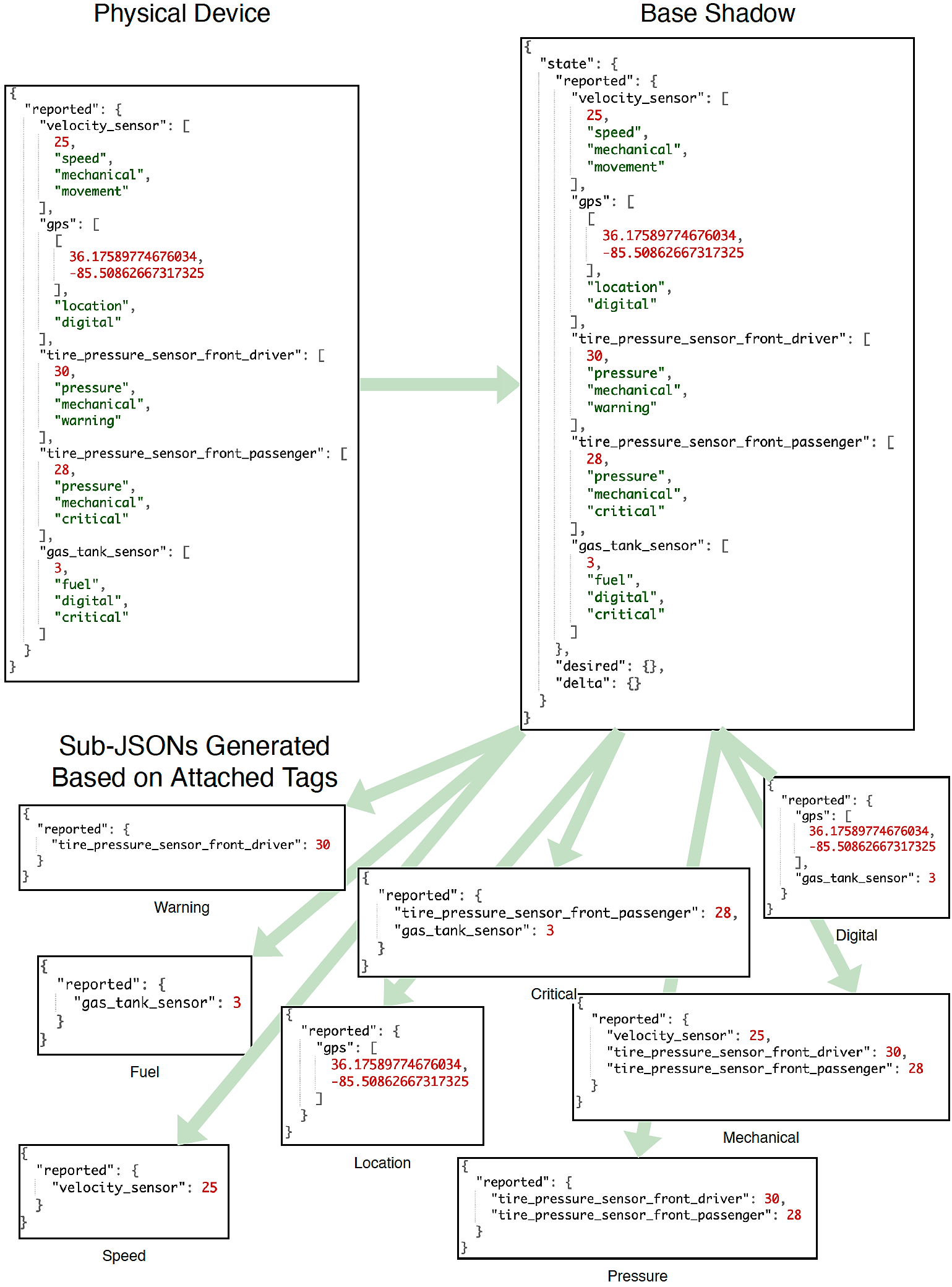}
    \caption{Propagation of Reported States to Sub-JSONs}
    \label{sub_json_generation}
\end{figure*}
Digital twins are the source of this subdivision, as they can exist in a many-to-one relationship with their physical counterparts. Each shadow instance holds a subset of the data present and can independently grant access to resources. These resources may query the shadow for the current system state, or publish desired states to the system. The resulting architecture leads to a distribution of data, and prevents a single MQTT client assuming all interaction with resources wishing to read or affect system state. The modularity of the separation of data into many separate digital twins also affords flexibility because not all clients must be active at any given point in time. Twins have the potential to be spun up or spun down as necessitated by resources in a form of load balancing. If a digital twin registers long periods of disconnection or inactivity from its associated device, the client could be halted until the device either has a state to report or the subset of information the client holds is requested by an external resource. This reactivity could be converted to a highly dynamic and scalable system which manages the number of active twins in real time based on demand. 

The implementation of physical devices is as straightforward in this architecture as it is in a one-to-one device-to-twin structure. Due to the centralized nature of MQTT, physical devices need only subscribe to topics following a pre-defined API (Application Programming Interface) structure to receive state change control signals. Authorization to publish to those topics may be handled by the broker, giving a central point at which access control can be done regarding all digital twins. This ensures security of the channels in which interaction takes while requiring few subscriptions from the physical device.

\subsection{Proposed Architecture}
The distribution and subdivision of data in our architecture is facilitated by the application of tags. Each key-value pair in the system holds a key string describing the meaning of the data held in the object and a value array containing the sensor value and tags attached to the object. These tags identify the function of that data within the implementation, provide structure for groupings of related data, and are the central mechanism for access control.
In this architecture tags support grouping of data by allowing similar data to be quickly associated and divided into subsets. Figure \ref{sub_json_generation} shows the processing of reported states (from the physical device to base shadow) with attached tags, and the division of data (from base shadow to multiple sub JSONs) based on those tags. For example in smart cars there are many different sets of data that could be produced such as speed, location, pressure, temperature, etc. All sensors in the car would then attach `pressure' to data measuring a pressure in the car. Additionally, more specific subsets can be made in order to grant external resources access to only the information they require. Therefore pressure data being monitored associated with the tires of the car may be tagged `tire' as well as `pressure' in order to differentiate it and allow more specific data sharing. Tags serve to group the data into most specific subset possible, after which the key associated denotes exactly what that data represents in the system.
 
Tags and key-value pairs hold a many-to-many relationship where one tag may be applied to many key-value pairs and conversely one key-value pair may hold many tags. This relationship allows data values to be distributed and held by many shadows, and also one shadow may receive many data values at once if a single tag is distributed to multiple key-value pairs in a message. As discussed earlier, where many key-value pairs are tagged `pressure' in a single message and subsequently distributed to the 'pressure' digital twin.
\begin{figure}[t!]
%\vspace{0.1in}
  %\centering
  \includegraphics[width=\linewidth]{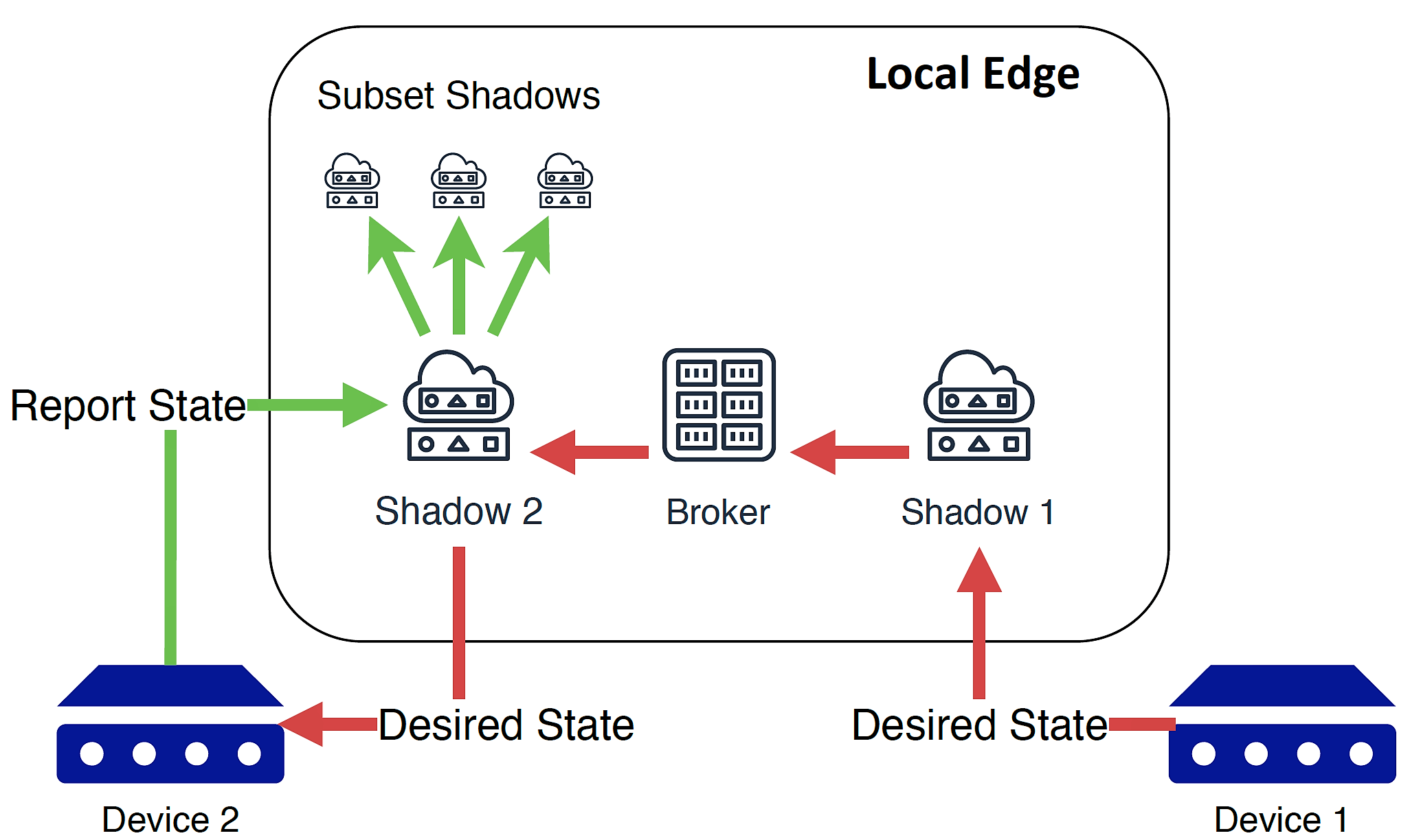}
  \caption{Implemented System Architecture}
  \label{des_state_resolution}
%\vspace{0.1in}
\end{figure}
\subsection{Assignment of Tags}
 The assignment of tags within a TBAC architecture must follow proper security practices, as assigned tags are the basis of access control. If tags are improperly assigned and therefore data is distributed to digital twins in which it should not reside, then resources that are given access to those twins will be served data they are not authorized to view. The attachment of low-security classification tags to high-security pieces of data is a simple way to gain access to critical data within the system. For example, if there exists a `timing' tag that functions as a benchmark to synchronize elements of the system then all resources would be able to access the digital twin containing `timing' information. If administrator is able to attach the `timing' tag to a piece of sensitive information such as location, or user data, it will enable unauthorized data read via the `timing' digital twin. 

 Tags should only be malleable to a few key authenticated resources in the system. They may be applied by the physical device itself based on characteristics of the data being generated. This is the foundation of on-the-fly dynamic tag attachment within the architecture. For example, a smart temperature sensor may have a ceiling at which the recorded temperature is no longer safe. When the recorded temperature exceeds that ceiling a `warning' tag could be applied to the data in order to trigger a safety system or inform the user. Additional levels may be present as well, so if the temperature exceeds another threshold a `critical' tag may be applied. Therefore response behavior can vary dependent upon the level of device failure. These tags can be attached when a value exceeds or falls below a predefined set-point and are device specific. The attachment of these tags means the associated key-value pair will be placed within the `warning' or `critical' digital twins, which allows all system health to be monitored via a small number of digital twins. This functionality is shown in Figure \ref{sub_json_generation}, where both tire pressure sensors are reporting values which have system health tags attached. The driver-side sensor has applied a `warning' tag which may be applied when the tire falls below the recommended specification by a relatively small margin. The passenger-side sensor has applied a `critical' tag which may be applied when the pressure falls too far e.g. below 30. 

 System administrators should also be able to add tags to device data as necessary. Therefore tags published to the device by the base unnamed shadow are applied to all further data generated by the device. Only the base shadow should have this functionality as it is the most controlled due to the centralized nature of the data it holds, and therefore administrators should be the only users with access.

\begin{table}[t!]
\begin{center}
\caption{Raspberry Pi 4 Model B Specifications}
\renewcommand{\arraystretch}{1.5}
 \begin{tabular}{ | p{0.60in} | p{2.25in} | } 
 \hline
 Operating System & Raspberry Pi OS, May 7, 2021 \\ 
 \hline
 CPU & Broadcom BCM2711, Quad core Cortex-A72 (ARM v8) 64-bit SoC @ 1.5GHz \\
 \hline
 RAM & 4 GB LPDDR4-3200 SDRAM \\
 \hline 
 Network Interface & Gigabit Ethernet, 2.4GHz and 5GHz 802.11b/g/n/ac Wi-Fi \\
 \hline
\end{tabular}
\label{tab:Raspberry-pi-specs}
\end{center}
\end{table}

\section{Implementation and Evaluation}
\label{sec:evaluation}

\begin{figure}[t!]
%\vspace{0.1in}
  \centering
  \includegraphics[width=7cm, height=3.4in]{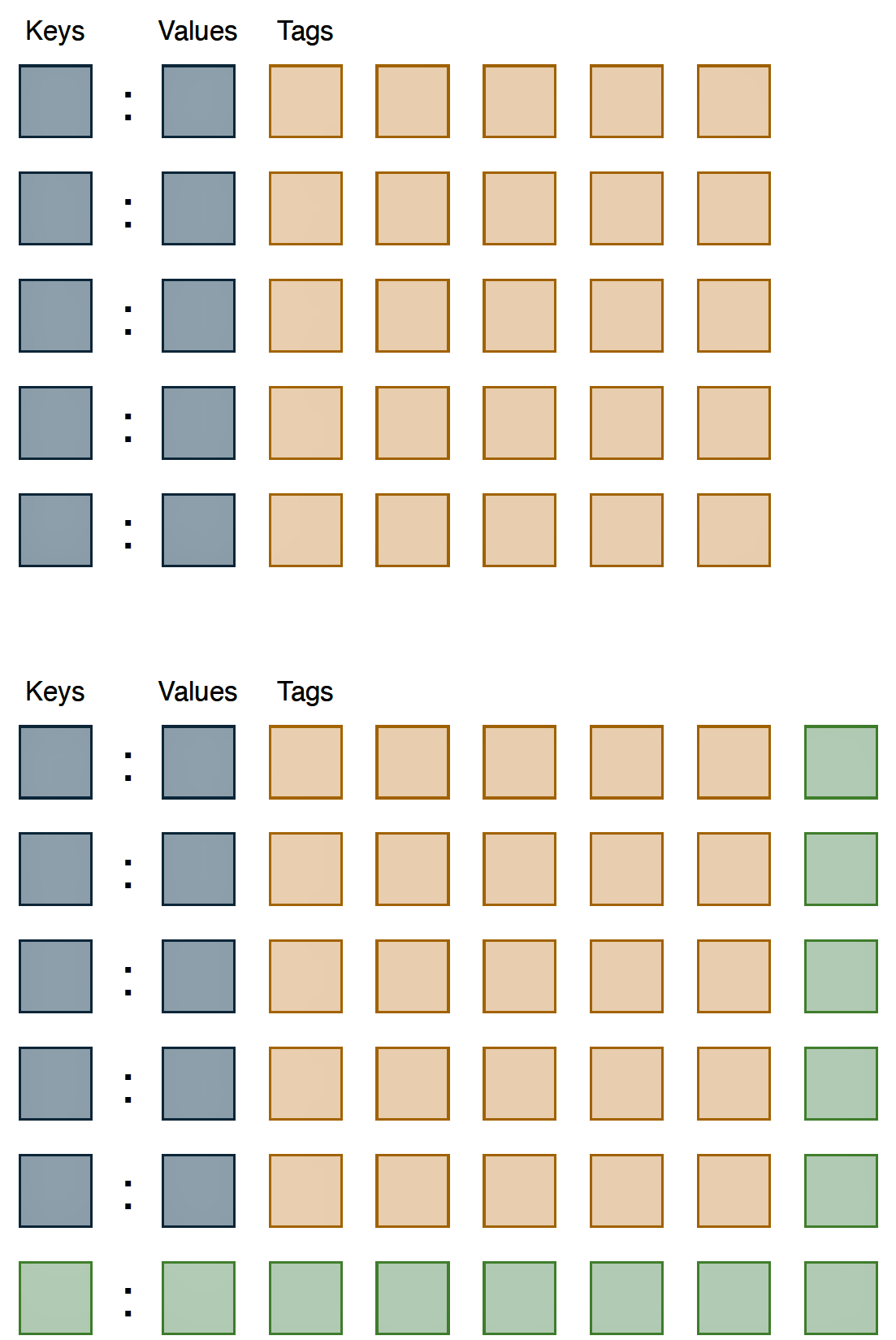}
  \caption{Transformation from Five to Six Key-Value Pairs}
  \label{experiment_state_transform}
%\vspace{0.1in}
\end{figure}
The proposed architecture has been evaluated using a lightweight digital twin implementation written in python and emulating AWS MQTT topic structure. This allows for large scalability and integration into industrial environments with minimal modification of the current code base. In this section, mechanism of interaction with devices in this architecture will be explained, then the performance of our implementation will be evaluated, finally application to industry will be discussed.

 Device state is divided into three subgroups: \textit{reported}, \textit{desired}, and \textit{delta}. The physical device modifies the reported subgroup when it is connected and reports its current state. Desired states may be pushed to the digital twin created in the local edge by authorized external clients in order to request a change in physical device state. If a difference is determined between the reported and desired states of the device, then the differing keys are added to the delta subgroup. When the device is connected, the calculated delta state is published to the physical device. Once the device receives these keys and transitions state, it reports the new state. Upon reception of a reported state matching a given desired state, the digital twin acknowledges that state as resolved and removes the associated key from both the desired and delta subgroups. The key-value pairs present in the reported state are then divided based on attached tags and distributed to their associated base shadows in the local edge, as shown in Fig. \ref{des_state_resolution}. The shadow and device python clients were run on a Raspberry Pi 4 for data collection, the specifications of which can be seen in Table \ref{tab:Raspberry-pi-specs}. The open source broker Mosquitto\footnote{https://mosquitto.org/} was used for authentication as it includes username/password, PSK (Pre-Shared Key), and external plugin support. The system was not stressed with the entirety of the architecture running on one device, and this should not affect the timing data collected as only the processing time of the digital twin is being evaluated.

\begin{figure}[t!]
%\vspace{0.1in}
  \centering
  \includegraphics[width=\linewidth, height=2.15in]{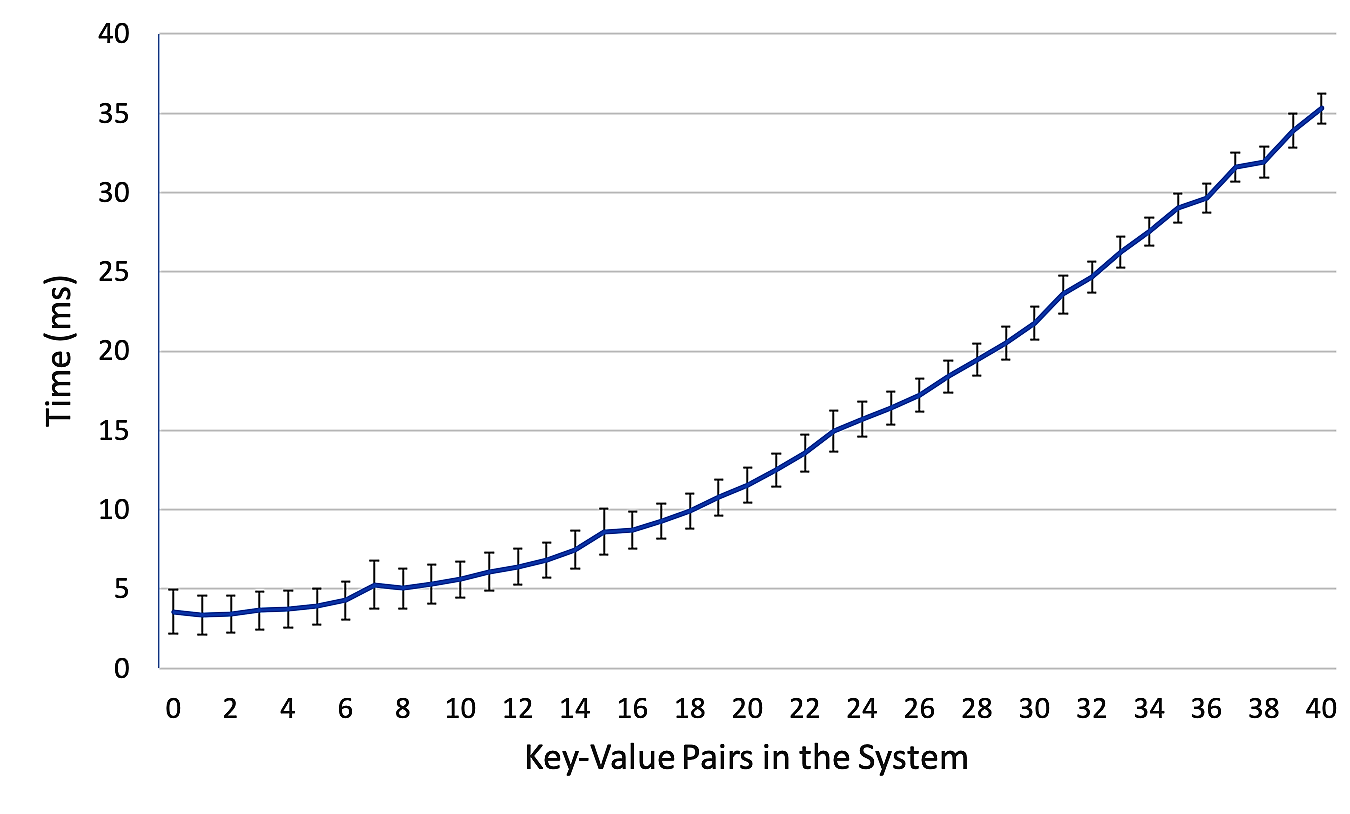}
  \caption{Average Dynamic Tag Assignment Processing Time}
  \label{dynamic_processing_time_graph}
%\vspace{0.1in}
\end{figure}
\begin{figure}[t!]
%\vspace{0.1in}
  \centering
  \includegraphics[width=\linewidth, height=2.5in]{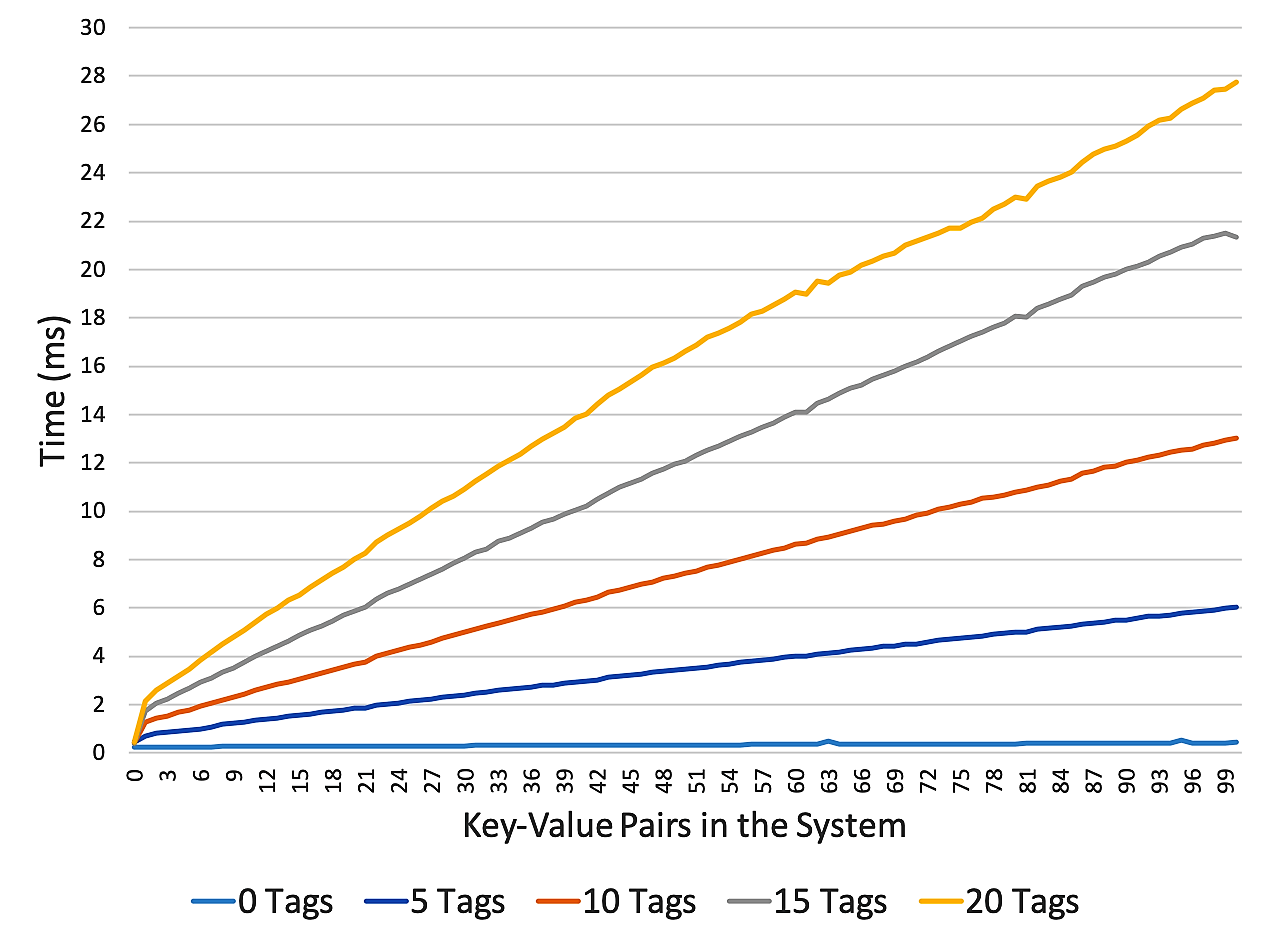}
  \caption{Average Static Tag Assignment Processing Time}
  \label{static_processing_time_graph}
%\vspace{0.1in}
\end{figure}

In order to evaluate the system's performance, the shadow linearly scales the relationship between the number of key-value pairs and tags associated with each pair in the system at a given point, as shown in Fig. \ref{experiment_state_transform}. For example, when there are five key-value pairs in the system there will be five tags attached to each of those pairs. Upon reception of a reported state from the simulated device, the shadow introduces a new key-value pair as well as increments the number of tags attached to all other pairs previously present. Therefore once the five pair state is reported by the device, a sixth pair is added and a new tag is appended to each of the five pairs already present. The new desired state of the system is then added to delta subgroup and subsequently published to simulated device. Once the simulated device conforms to the received desired state and reports its current state, this cycle continues. 
To effectively measure the efficiency of our architecture, we calculate time only while the shadow is resolving messages. This avoids introduction of variance from network delay as well as abstracts our evaluation away from device specific implementations. Different IoT devices will have widely varying computational hardware as well as polling rates depending on implementation specific variables. The exclusion of device interaction from collected timings leads to more consistent results and a stronger examination of our implementation of the proposed architecture.

There are three distinct function calls included in the timings collected: `update', `delta', and `parse\_tags'. Update function processes the incoming message and places relevant key-value pairs into their designated position within the JSON structure. The delta function balances the JSON structure to ensure it retains continuity and consistency regarding the AWS-style. This encompasses functionality discussed earlier, such as removing keys from desired subgroup once a matching reported state has been received. Once the JSON has been cleaned the remaining keys in delta subgroup are published to device. The final call is to `parse\_tags', which compiles a list of all tags attached to key-value pairs in the message and generates and publishes sub-JSONs to associated shadows.

Figure \ref{dynamic_processing_time_graph} shows the processing time of the system averaged over 500 trials where each trial incremented the number of key-value pairs in the system from zero to forty and then emptied the system. Each point is the average performance time of the system associated with that many key-value pairs present. The attached error bars indicate a 99\% confidence interval and show that with 1600 total tags present in the system (40 attached to 40 key-value pairs), the average processing time will rarely exceed 36 ms.
 Figure \ref{static_processing_time_graph} displays the variation in system function with a static number of tags attached to each key-value pair present. Each data series represents an incrementation of number of key-value pairs from zero to one hundred with each additional pair containing specified number of tags, e.g. at data point 80 in the `5 Tags' series there are 80 key-value pairs each of which has 5 attached tags. Each data series is the average of 500 trials where each trial represents the filling of the system from zero to one hundred pairs.

These results are promising for real time and edge centric industry applications as they show minimal increase in processing time at the digital twin level, while largely scaling the number of tags and data pairs present in the system. Due to the exponential nature of the `parse\_tags' function, it is the largest bottleneck of the system. However, if the number of tags applied to each key-value pair is kept low the system remains scalable and suffers minor performative degradation.

\section{Conclusion}
\label{sec:conclusion}
This research demonstrates edge centric access control structure in IoT environments by proposing a novel TBAC architecture focused on the division of data into multiple digital twins. This architecture fills a gap in environments where on the fly and real time limited data exposure is highly critical, and allows for complete subdivision of data based on tags attached directly to present data. Complex device relationships are supported via the many-to-many relationship between tags and data, allowing implementations to model peculiar environments with little additional complexity. We discuss the usefulness of this architecture in smart environments such as manufacturing and internet-connected vehicles, and give an example of the flow of tagged data in these environment. Industry solutions currently offered have been examined regarding their integration of TBAC as well as their capacity to divide data into subsets. The weaknesses and strengths of offered services are discussed in relation to the proposed architecture. We deployed a local implementation of our architecture and examined the effect of number of attached tags on performance. We envision further exploration regarding access control on the tagged shadows, and the application of this data distribution to other smart environments.

\section*{Acknowledgement}
%This work was supported in part by the NSF at TTU under grant 2025682, and in part by the NSF CREST Center at UTSA under Grant HRD-1736209.
This research is supported by NSF CREST Center Grant
HRD-1736209 at UTSA, and by the Grant 2025682 at TTU.

\bibliographystyle{IEEEtran}
\bibliography{main}

\begin{thebibliography}{10}
\providecommand{\url}[1]{#1}
\csname url@samestyle\endcsname
\providecommand{\newblock}{\relax}
\providecommand{\bibinfo}[2]{#2}
\providecommand{\BIBentrySTDinterwordspacing}{\spaceskip=0pt\relax}
\providecommand{\BIBentryALTinterwordstretchfactor}{4}
\providecommand{\BIBentryALTinterwordspacing}{\spaceskip=\fontdimen2\font plus
\BIBentryALTinterwordstretchfactor\fontdimen3\font minus
  \fontdimen4\font\relax}
\providecommand{\BIBforeignlanguage}[2]{{%
\expandafter\ifx\csname l@#1\endcsname\relax
\typeout{** WARNING: IEEEtran.bst: No hyphenation pattern has been}%
\typeout{** loaded for the language `#1'. Using the pattern for}%
\typeout{** the default language instead.}%
\else
\language=\csname l@#1\endcsname
\fi
#2}}
\providecommand{\BIBdecl}{\relax}
\BIBdecl

\bibitem{205156}
E.~Zeng \emph{et~al.}, ``End user security and privacy concerns with smart
  homes,'' in \emph{Thirteenth Symposium on Usable Privacy and Security}.

\bibitem{holst_2021}
\BIBentryALTinterwordspacing
A.~Holst, ``{IoT connected devices worldwide 2019-2030},'' Jan 2021. [Online].
  Available:
  \url{https://www.statista.com/statistics/1183457/iot-connected-devices-worldwide/}
\BIBentrySTDinterwordspacing

\bibitem{onestore}
\BIBentryALTinterwordspacing
``{Nokia: Threat Intelligence Report 2020}.'' [Online]. Available:
  \url{https://onestore.nokia.com/asset/210088}
\BIBentrySTDinterwordspacing

\bibitem{nam}
\BIBentryALTinterwordspacing
``{2019 United States Manufacturing Facts}.'' [Online]. Available:
  \url{https://www.nam.org/state-manufacturing-data/2019-united-states-manufacturing-facts/}
\BIBentrySTDinterwordspacing

\bibitem{wagner_2020}
\BIBentryALTinterwordspacing
I.~Wagner, ``Worldwide - connected car shipments,'' Sep 2020. [Online].
  Available:
  \url{https://www.statista.com/statistics/743400/estimated-connected-car-shipments-globally/}
\BIBentrySTDinterwordspacing

\bibitem{digitaltwins}
\BIBentryALTinterwordspacing
M.~Batty, ``Digital twins,'' \emph{Environment and Planning B: Urban Analytics
  and City Science}, vol.~5, 2018. [Online]. Available:
  \url{https://doi.org/10.1177/2399808318796416}
\BIBentrySTDinterwordspacing

\bibitem{7809752}
A.~Alshehri and R.~Sandhu, ``Access control models for cloud-enabled internet
  of things: A proposed architecture and research agenda,'' in \emph{IEEE CIC},
  2016, pp. 530--538.

\bibitem{gupta2018authorization}
M.~Gupta and R.~Sandhu, ``Authorization framework for secure cloud assisted
  connected cars and vehicular internet of things,'' in \emph{Proc. of the ACM
  SACMAT}, 2018, pp. 193--204.

\bibitem{Alshehri2018AccessCM}
A.~Alshehri \emph{et~al.}, ``Access control model for virtual objects (shadows)
  communication for aws internet of things,'' \emph{Proc. of the ACM CODASPY},
  2018.

\bibitem{8673782}
K.-H.~N. Bui and J.~J. Jung, ``Aco-based dynamic decision making for connected
  vehicles in iot system,'' \emph{IEEE Transactions on Industrial Informatics},
  vol.~15, no.~10, pp. 5648--5655, 2019.

\bibitem{weijia2018}
W.~He \emph{et~al.}, ``Rethinking access control and authentication for the
  home internet of things (iot),'' in \emph{27th {USENIX} Security Symposium},
  2018.

\bibitem{weijia2021}
H.~Weijia, ``Sok: Context sensing for access control in the adversarial home
  iot.'' 2021.

\bibitem{userauthIoT}
Y.~Tian \emph{et~al.}, ``{SmartAuth: User-Centered Authorization for the
  Internet of Things},'' in \emph{USENIX Security Symposium}, 2017, pp.
  361--378.

\bibitem{celik_tan_mcdaniel_2019}
Z.~B. Celik, G.~Tan, and P.~McDaniel, ``{Iotguard: Dynamic enforcement of
  security and safety policy in commodity iot},'' \emph{Proc. of NDSS}, 2019.

\bibitem{franch2020}
M.~Fernández \emph{et~al.}, ``A data access model for privacy-preserving
  cloud-iot architectures,'' \emph{Proc. of the ACM SACMAT}, 2020.

\bibitem{yahyazadeh2019}
M.~Yahyazadeh \emph{et~al.}, ``Expat: Expectation-based policy analysis and
  enforcement for appified smart-home platforms,'' in \emph{Proc. of the ACM
  Symposium on Access Control Models and Technologies}, 2019.

\bibitem{yahyazadeh2020}
------, ``Patriot: Policy assisted resilient programmable iot system,'' in
  \emph{International Conference on Runtime Verification}.\hskip 1em plus 0.5em
  minus 0.4em\relax Springer, 2020.

\bibitem{gupta2020attribute}
M.~Gupta \emph{et~al.}, ``An attribute-based access control for cloud enabled
  industrial smart vehicles,'' \emph{IEEE Transactions on Industrial
  Informatics}, vol.~17, no.~6, pp. 4288--4297, 2020.

\bibitem{dgupta2020access}
D.~Gupta \emph{et~al.}, ``{Access control model for Google Cloud IoT},'' in
  \emph{IEEE Conf. on Big Data Security on Cloud}, 2020, pp. 198--208.

\bibitem{guptaABAC2019}
M.~Gupta \emph{et~al.}, ``Dynamic groups and attribute-based access control for
  next-generation smart cars,'' in \emph{Proc. of the ACM CODASPY}, 2019.

\bibitem{gupta2020secure}
------, ``{Secure V2V and V2I communication in intelligent transportation using
  cloudlets},'' \emph{IEEE Transactions on Services Computing}, 2020.

\bibitem{9502070}
S.~Bhatt \emph{et~al.}, ``{Attribute-Based Access Control for AWS Internet of
  Things and Secure Industries of the Future},'' \emph{IEEE Access}, 2021.

\bibitem{gupta2020security}
M.~Gupta \emph{et~al.}, ``Security and privacy in smart farming: Challenges and
  opportunities,'' \emph{IEEE Access}, vol.~8, pp. 34\,564--34\,584, 2020.

\bibitem{sina2020farming}
S.~Sontowski \emph{et~al.}, ``Cyber attacks on smart farming infrastructure,''
  in \emph{IEEE Int. Conf. on Collaboration and Internet Computing}, 2020.

\bibitem{o2016insecurity}
M.~O’Neill \emph{et~al.}, ``Insecurity by design: Today’s iot device
  security problem,'' \emph{Engineering}, vol.~2, no.~1, pp. 48--49, 2016.

\bibitem{dhanda_singh_jindal_2020}
S.~S. Dhanda \emph{et~al.}, ``Lightweight cryptography: A solution to secure
  iot,'' \emph{Wireless Personal Comm.}, vol. 112, no.~3, p. 1947–1980, 2020.

\bibitem{kusiak2018smart}
A.~Kusiak, ``Smart manufacturing,'' \emph{International Journal of Production
  Research}, vol.~56, no. 1-2, pp. 508--517, 2018.

\bibitem{s20195480}
P.~Trakadas \emph{et~al.}, ``An artificial intelligence-based collaboration
  approach in industrial iot manufacturing: Key concepts, architectural
  extensions and potential applications,'' \emph{Sensors}, vol.~20, no.~19,
  2020.

\end{thebibliography}

\end{document}